\documentclass[a4paper]{jpconf}
\usepackage{kpfonts}
\usepackage{graphicx}
\usepackage{xspace}
\newcommand{\mjd}{\textsc{Majorana Demonstrator}\xspace}
\newcommand{\mj}{\textsc{Majorana}\xspace}
\newcommand{\dem}{\textsc{Demonstrator}\xspace}
\newcommand{\znbb}{$0\nu\beta\beta$\xspace}
\newcommand{\tnbb}{$2\nu\beta\beta$\xspace}
\newcommand{\gei}{$^{76}$Ge\xspace}

\newcommand{\unc}{$^{1}$}
\newcommand{\tunl}{$^{2}$}
\newcommand{\lbnl}{$^{3}$}
\newcommand{\uw}{$^{4}$}
\newcommand{\pnnl}{$^{5}$}
\newcommand{\usc}{$^{6}$}
\newcommand{\ornl}{$^{7}$}
\newcommand{\itep}{$^{8}$}
\newcommand{\usd}{$^{9}$}
\newcommand{\mpi}{$^{10}$}
\newcommand{\sdsmt}{$^{11}$}
\newcommand{\jinr}{$^{12}$}
\newcommand{\duke}{$^{13}$}
\newcommand{\lanl}{$^{14}$}
\newcommand{\ut}{$^{15}$}
\newcommand{\ou}{$^{16}$}
\newcommand{\princeton}{$^{17}$}
\newcommand{\ncsu}{$^{18}$}
\newcommand{\blhill}{$^{19}$}
\newcommand{\ttu}{$^{20}$}
\newcommand{\queens}{$^{21}$}
\newcommand{\tum}{$^{22}$}
\newcommand{\cs}{$^{,}$}
\newcommand{\clara}{$^{a}$}

\begin{document}

\title{Initial results from the \mjd}

\author{T~S~Caldwell\unc\cs\tunl, N~Abgrall\lbnl, S~I~Alvis\uw, I~J~Arnquist\pnnl, F~T~Avignone~III\usc\cs\ornl, A~S~Barabash\itep, C~J~Barton\usd, F~E~Bertrand\ornl, T~Bode\mpi, B~Bos\sdsmt, A~W~Bradley\lbnl, V~Brudanin\jinr, M~Busch\duke\cs\tunl, M~Buuck\uw, Y-D~Chan\lbnl, C~D~Christofferson\sdsmt, P~-H~Chu\lanl, C~Cuesta\uw\cs\clara, J~A~Detwiler\uw, C~Dunagan\sdsmt, Yu~Efremenko\ut\cs\ornl, H~Ejiri\ou, S~R~Elliott\lanl, T~Gilliss\unc\cs\tunl, G~K~Giovanetti\princeton, M~P~Green\ncsu\cs\tunl\cs\ornl, J~Gruszko\uw, I~S~Guinn\uw, V~E~Guiseppe\usc, C~R~Haufe\unc\cs\tunl, L~Hehn\lbnl, R~Henning\unc\cs\tunl, E~W~Hoppe\pnnl, M~A~Howe\unc\cs\tunl, K~J~Keeter\blhill, M~F~Kidd\ttu, S~I~Konovalov\itep, R~T~Kouzes\pnnl, A~M~Lopez\ut, R~D~Martin\queens, R~Massarczyk\lanl, S~J~Meijer\unc\cs\tunl, S~Mertens\mpi\cs\tum, J~Myslik\lbnl, C~O'Shaughnessy\unc\cs\tunl, G~Othman\unc\cs\tunl, W~Pettus\uw, A~W~P~Poon\lbnl, D~C~Radford\ornl, J~Rager\unc\cs\tunl, A~L~Reine\unc\cs\tunl, K~Rielage\lanl, R~G~H~Robertson\uw, N~W~Ruof\uw, B~Shanks\unc\cs\tunl, M~Shirchenko\jinr, A~M~Suriano\sdsmt, D~Tedeschi\usc, J~E~Trimble\unc\cs\tunl, R~L~Varner\ornl, S~Vasilyev\jinr, K~Vetter\lbnl, K~Vorren\unc\cs\tunl, B~R~White\lanl, J~F~Wilkerson\unc\cs\tunl\cs\ornl, C~Wiseman\usc, W~Xu\usd, E~Yakushev\jinr, C~-H~Yu\ornl, V~Yumatov\itep, I~Zhitnikov\jinr, and B~X~Zhu\lanl}                                                          
\address{\unc Department of Physics and Astronomy, University of North Carolina, Chapel Hill, NC, USA}
\address{\tunl Triangle Universities Nuclear Laboratory, Durham, NC, USA}
\address{\lbnl Nuclear Science Division, Lawrence Berkeley National Laboratory, Berkeley, CA, USA}
\address{\uw Center for Experimental Nuclear Physics and Astrophysics, and Department of Physics, University of Washington, Seattle, WA, USA}
\address{\pnnl Pacific Northwest National Laboratory, Richland, WA, USA}
\address{\usc Department of Physics and Astronomy, University of South Carolina, Columbia, SC, USA}
\address{\ornl Oak Ridge National Laboratory, Oak Ridge, TN, USA}
\address{\itep National Research Center ``Kurchatov Institute'' Institute for Theoretical and Experimental Physics, Moscow, Russia}
\address{\usd Department of Physics, University of South Dakota, Vermillion, SD, USA} 
\address{\mpi Max-Planck-Institut f\"{u}r Physik, M\"{u}nchen, Germany}
\address{\sdsmt South Dakota School of Mines and Technology, Rapid City, SD, USA}
\address{\jinr Joint Institute for Nuclear Research, Dubna, Russia}
\address{\duke Department of Physics, Duke University, Durham, NC, USA}
\address{\lanl Los Alamos National Laboratory, Los Alamos, NM, USA}
\address{\ut Department of Physics and Astronomy, University of Tennessee, Knoxville, TN, USA}
\address{\ou Research Center for Nuclear Physics, Osaka University, Ibaraki, Osaka, Japan}
\address{\princeton Department of Physics, Princeton University, Princeton, NJ, USA}
\address{\ncsu Department of Physics, North Carolina State University, Raleigh, NC, USA}
\address{\blhill Department of Physics, Black Hills State University, Spearfish, SD, USA}
\address{\ttu Tennessee Tech University, Cookeville, TN, USA}
\address{\queens Department of Physics, Engineering Physics and Astronomy, Queen's University, Kingston, ON, Canada} 
\address{\tum Physik Department, Technische Universit\"{a}t, M\"{u}nchen, Germany}
\vspace{0.2cm}
\address{\clara Present Address: Centro de Investigaciones Energ\'{e}ticas, Medioambientales y Tecnol\'{o}gicas, CIEMAT, 28040, Madrid, Spain}
\vspace{0.2cm}
\ead{tcald@unc.edu}

\begin{abstract}
The \mj Collaboration has assembled an array of high purity Ge detectors to search for neutrinoless double-beta decay in \gei with the goal of establishing the required background and scalability of a Ge-based next-generation ton-scale experiment. The \mjd consists of 44~kg of high-purity Ge (HPGe) detectors (30~kg enriched in \gei) with a low-noise p-type point contact (PPC) geometry. The detectors are split between two modules which are contained in a single lead and high-purity copper shield at the Sanford Underground Research Facility in Lead, South Dakota. Following a commissioning run that started in June 2015, the full detector array has been acquiring data since August 2016. We will discuss the status of the \mjd and initial results from the first physics run; including current background estimates, exotic low-energy physics searches, projections on the physics reach of the \dem, and implications for a ton-scale Ge-based neutrinoless double-beta decay search.
\end{abstract}

\section{Introduction}
\label{sec:intro}

Searches for neutrinoless double-beta decay (\znbb) are the most direct experimental test of the Majorana nature of the neutrino~\cite{0nbb}.  Observation of \znbb decay would immediately imply that neutrinos are Majorana particles and that total lepton number is violated, having profound implications for physics beyond the Standard Model.  Developments in germanium detector technology have motivated searches for \znbb decay in \gei.  With complimentary efforts ongoing from the GERDA experiment~\cite{gerda}, the \mj collaboration is now operating an array of Ge detectors in the \dem with the primary goal of demonstrating the background and energy resolution required to justify a ton-scale \gei \znbb experiment capable of discovery level sensivity in the inverted neutrino mass ordering range.

\section{Experimental overview}
\label{sec:overview}

The \mjd consists of an array of 58 high-purity Ge detectors with a combined mass of 44.1~kg.  Isotopically enriched Ge was used to fabricate 35 of the detectors (29.7~kg) which have 88\% \gei content.  The remaining 14.4~kg of the detectors are of natural \gei abundance (7.8\%).  In order to minimize cosmogenic activation of the enriched Ge, the enriched material was stored underground and shielded through all phases of the detector fabrication~\cite{geprocessing}.  The detectors are P-type, point-contact (PPC)~\cite{ppc0, ppc1}, offering excellent energy resolution, pulse shape based background rejection, and low energy thresholds.

Wherever possible, components required for mounting and housing the detectors are made from underground electroformed copper (UGEFCu) that was machined in a dedicated underground machine shop.  Assay of the UGEFCu resulted in limits of $\leq0.1$~$\mu$Bq/kg $^{232}$Th and $\leq0.1$~$\mu$Bq/kg $^{238}$U~\cite{ugefcu}.  All other components near the detectors are selected based on a rigorous radio-assay program~\cite{ugefcu}.

\begin{figure}[h]
  \begin{center}
    \includegraphics[width=0.45\textwidth]{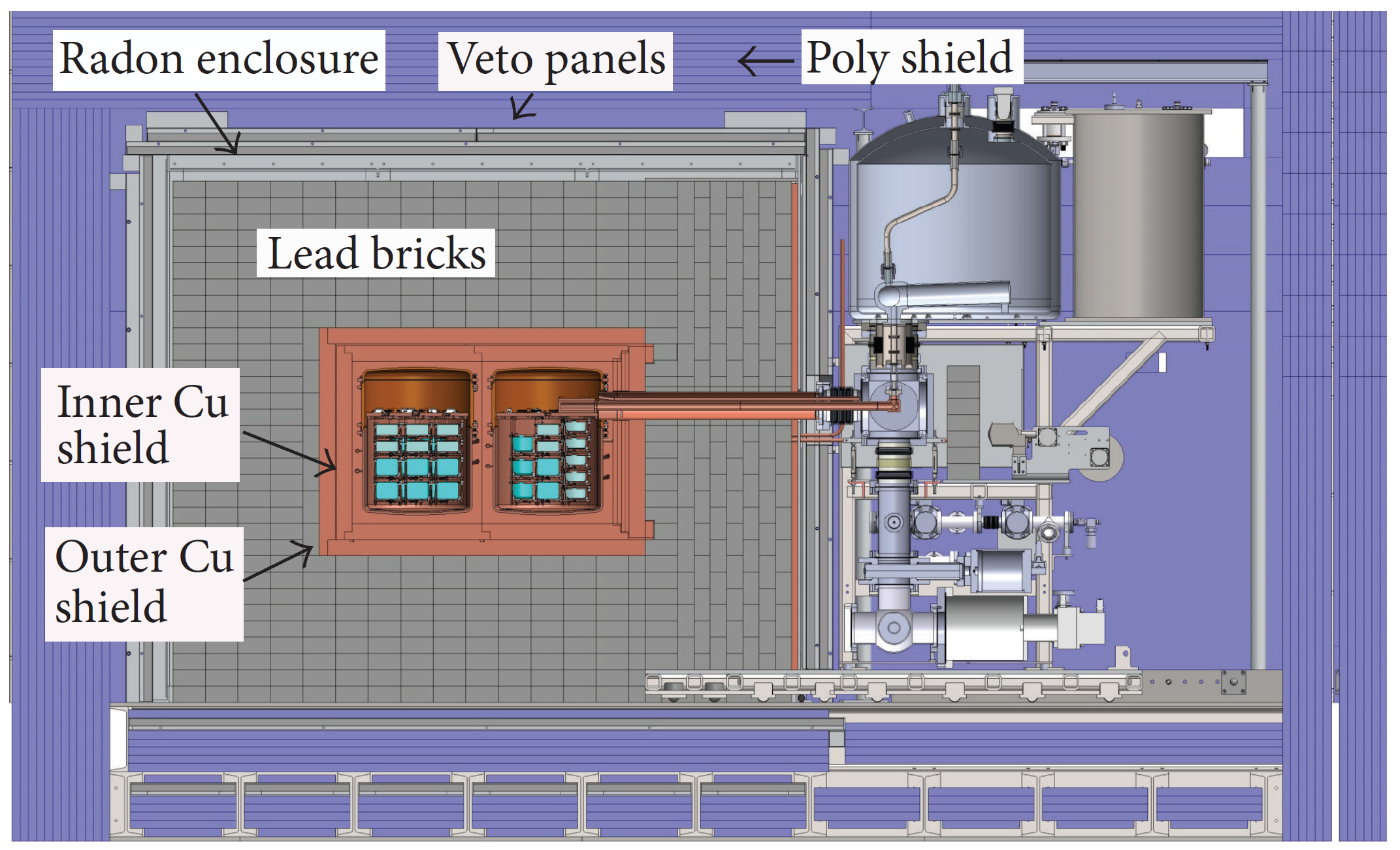}
    \caption{A cross-section of the \dem shield shown with both cryostats installed~\cite{mjdoverview}.  \label{fig:mjd_config}}%The vacuum and cryogenic system for the cryostat on the right is visible outside the lead shield, but within the polyethylene shielding. \label{fig:mjd_config}}
  \end{center}
\end{figure}

The detectors are split between two modules, each housed within an independent vacuum cryostat made from UGEFCu.  As shown in Figure~\ref{fig:mjd_config}, the cryostats are surrounded on all sides by a graded shield consisting of an inner layer of UGEFCu (5~cm), an outer layer of commercial ultra-pure copper (5~cm), high-purity lead (45~cm), an active muon veto, and high density polyethylene (30~cm).  The entire lead shield is contained in a sealed aluminum enclosure that is purged with liquid nitrogen boil-off to minimize the radon concentration near the cryostats.  Shielding from cosmogenics is achieved by operating the \dem at the 4850-foot level of the Sanford Underground Research Facility (SURF) in Lead, SD~\cite{surf}.

\section{Operations and data acquired}
\label{sec:ops}

Each configuration and operational state of the \mjd is assigned a Data Set (DS) beginning with DS0 (June-October 2015).  In this configuration, only Module 1 (M1) was installed in the shield, and the inner UGEFCu shield was not in place.  The backgrounds in M1 were reduced in DS1 (December 2015-May 2016) with installation of the UGEFCu shield.  In DS2 (May-July 2016), the digitizers were operated in a mode that pre-sums the waveforms after the rising edge to test possible improvements in the discrimination of alpha events.  Both modules were operated simultaneously in the shield during DS3 and DS4 (August-September 2016) with independent DAQ systems for M1 and M2 respectively.  In DS5 (October 2016-May 2017), the DAQ systems were merged into a single data stream, and blind data was acquired from March 17-May 11.  Currently the \dem is acquiring blind data in DS6 with the combined DAQ system and pre-summing of the post-rising edge waveforms.

\section{Background rejection and initial results}
\label{sec:results}

The properties of \mjd's PPC detectors offer additional background suppression beyond that provided by clean materials and a high purity shield.  The excellent energy resolution of PPC detectors allows a narrow \znbb region that minimizes background from the \tnbb spectrum and other continuum backgrounds.  In DS3 and DS4, the \dem has achieved a 2.4~keV FWHM at the $Q$-value of 2039~keV, the best energy resolution in the \znbb region of any double-beta decay experiment to date.

PPC detectors also have slow drift velocities throughout much of the crystal bulk with a highly localized weighting potential near the point contact.  The difference in drift times for energy depositions at different locations in the crystal allows multi-site events to be identified in contrast to single-site events like double-beta decays, which are single-site due to the short range of the emitted electrons.  In the \dem, the maximum of the current pulse (A) as a function of the reconstructed energy (E) is scaled using calibration data~\cite{calibration} to give 90\% acceptance of double-escape events from the $^{208}$Tl 2614~keV gamma.  This provides an `AvsE' parameter which is used to discriminate single-site from multi-site events~\cite{avse}.  %The $^{208}$Tl single-escape peak from calibrations provides a sample of multi-site events that are rejected with 94\% efficiency.

The lithiated dead layer covering much of the detector surface makes the detectors largely insensitive to alphas emitted by nearby contaminants.  However, alphas incident on the passivated surface surrounding the point-contact result in an energy-degraded signal with a component of the charge drifting slowly along the passivated surface.  This component with slow charge collection results in a characteristic slope to the waveform tail which can be used to identify alphas.  The delayed charge recovery (DCR) parameter scales the slope of the waveform tail to give 90\% acceptance (for the results shown here) of single-site events in the Compton shoulder of the 2614~keV $^{208}$Tl line from calibration data~\cite{dcr}.

\begin{figure}[h]
  \begin{center}
    \includegraphics[width=0.5\textwidth]{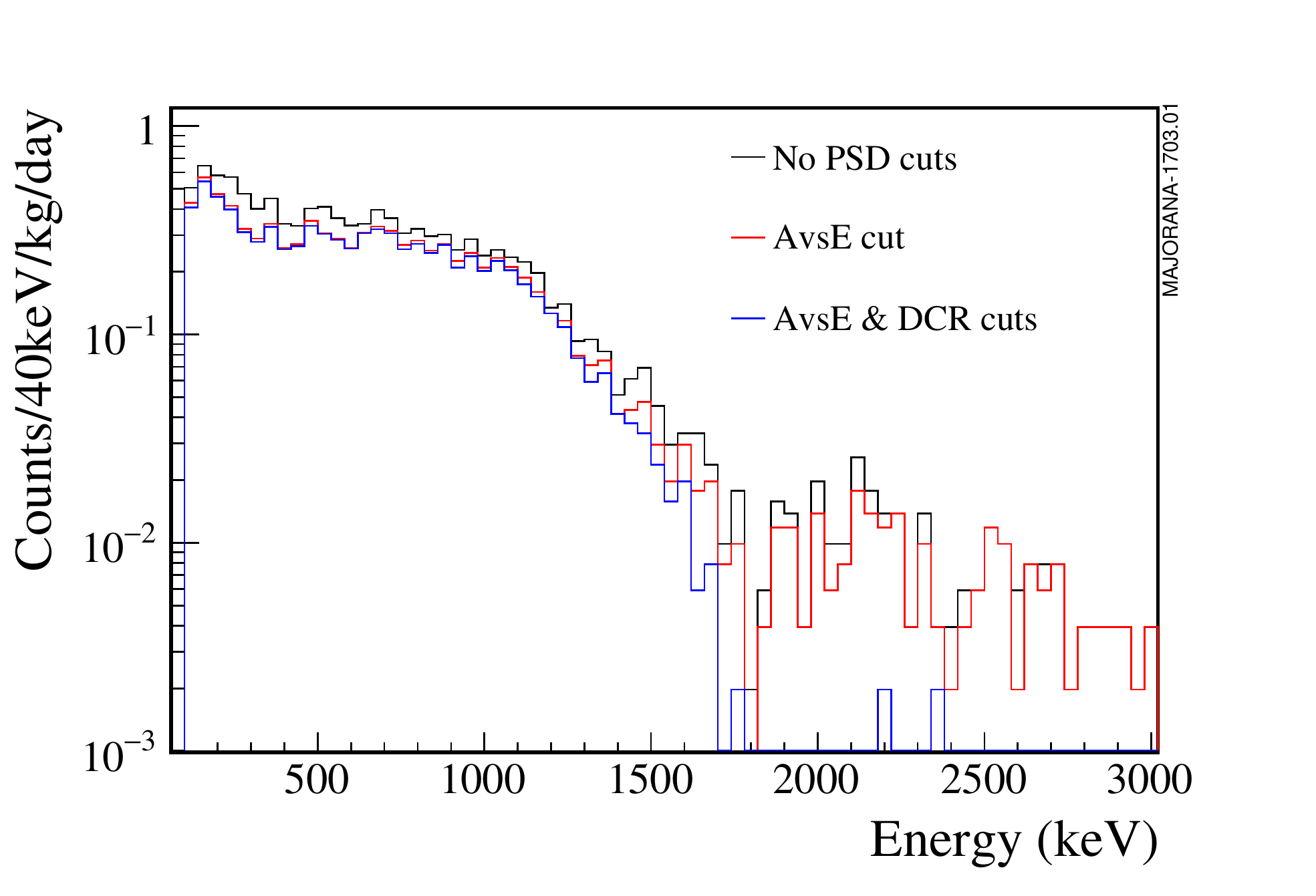}
    \caption{The enriched detector background spectrum from DS3 and DS4 with combined 1.39~kg-y of exposure and PSD cuts applied. \label{fig:ds34_spec}}%with AvsE~\cite{avse} and DCR~\cite{dcr} cuts applied to remove multi-site events and passivated surface alphas respectively.  \label{fig:ds34_spec}}
  \end{center}
\end{figure}

Before applying the pulse shape discrimination (PSD) cuts described above, instrumental backgrounds~\cite{datacleaning} and events with energy deposition in multiple detectors within a 4~$\mu$s coincidence window are removed.  The spectrum shown in Figure~\ref{fig:ds34_spec} is the result of the combination of DS3 and DS4 and sequentially-applied AvsE and DCR cuts with an exposure of 1.39 kg-y.  After all cuts, the \tnbb spectrum is the only visible feature.  In order to estimate the background index in the narrow \znbb region of interest (ROI), the background in a much wider 400~keV window centered at 2039~keV is used, and the background is scaled to the width of the ROI.  Assuming a Gaussian line shape, the optimal ROI width is 2.9~keV and 2.6~keV for M1 and M2 respectively.  This results in a projected background rate of $5.1^{+8.9}_{-3.2}$~c/ROI/t/y.  The corresponding background index is $1.8\times10^{-3}$~c/keV/kg/y.  Analysis of all data sets with a combined background is underway~\cite{specanalysis}, and a first \znbb limit from the \mjd is expected to soon be released.

\begin{figure}[h]
  \begin{center}
    $\begin{array}{c@{\hspace{0.25cm}}c}
      \includegraphics[width=0.425\textwidth]{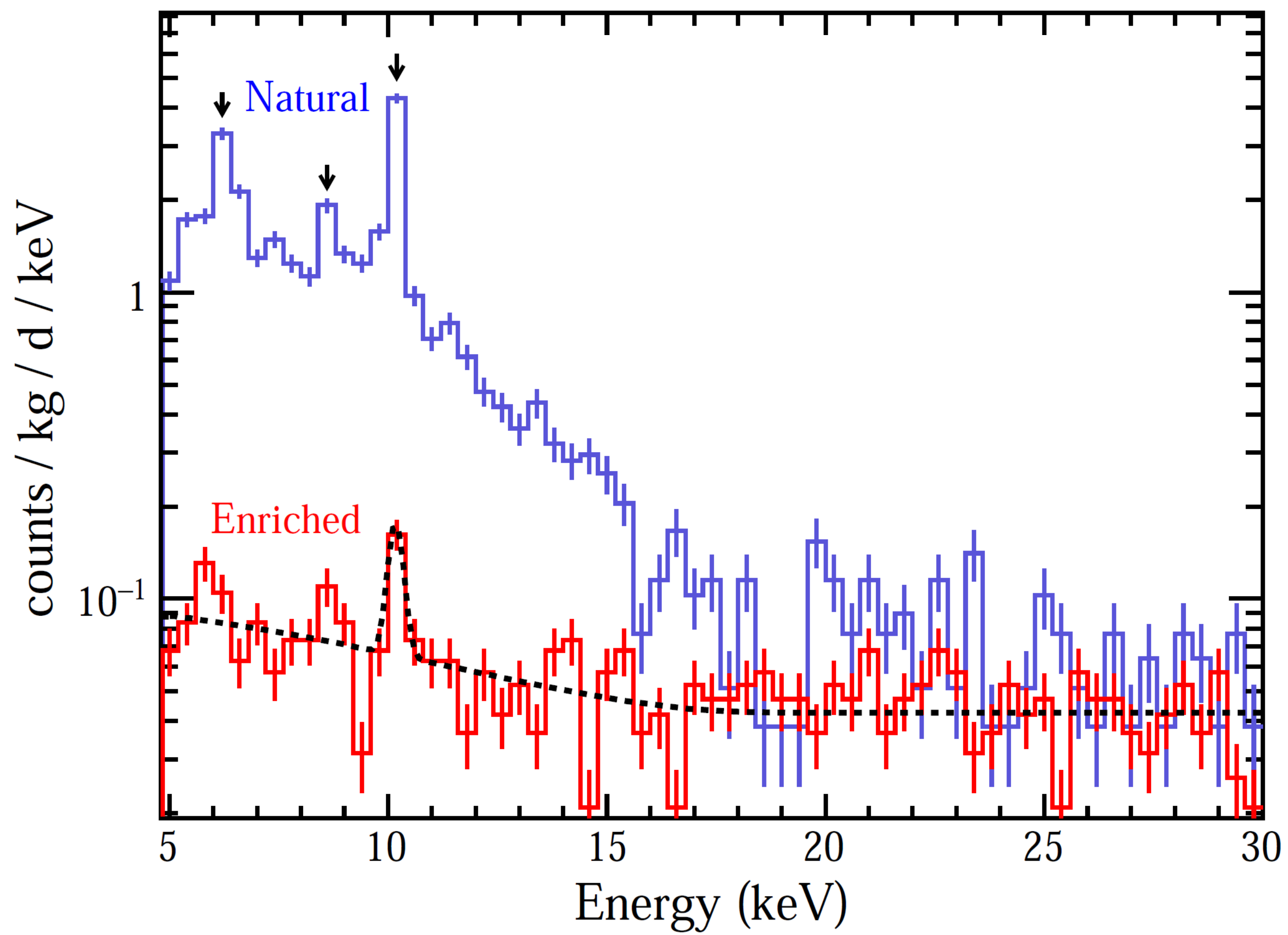} &
      \includegraphics[width=0.475\textwidth]{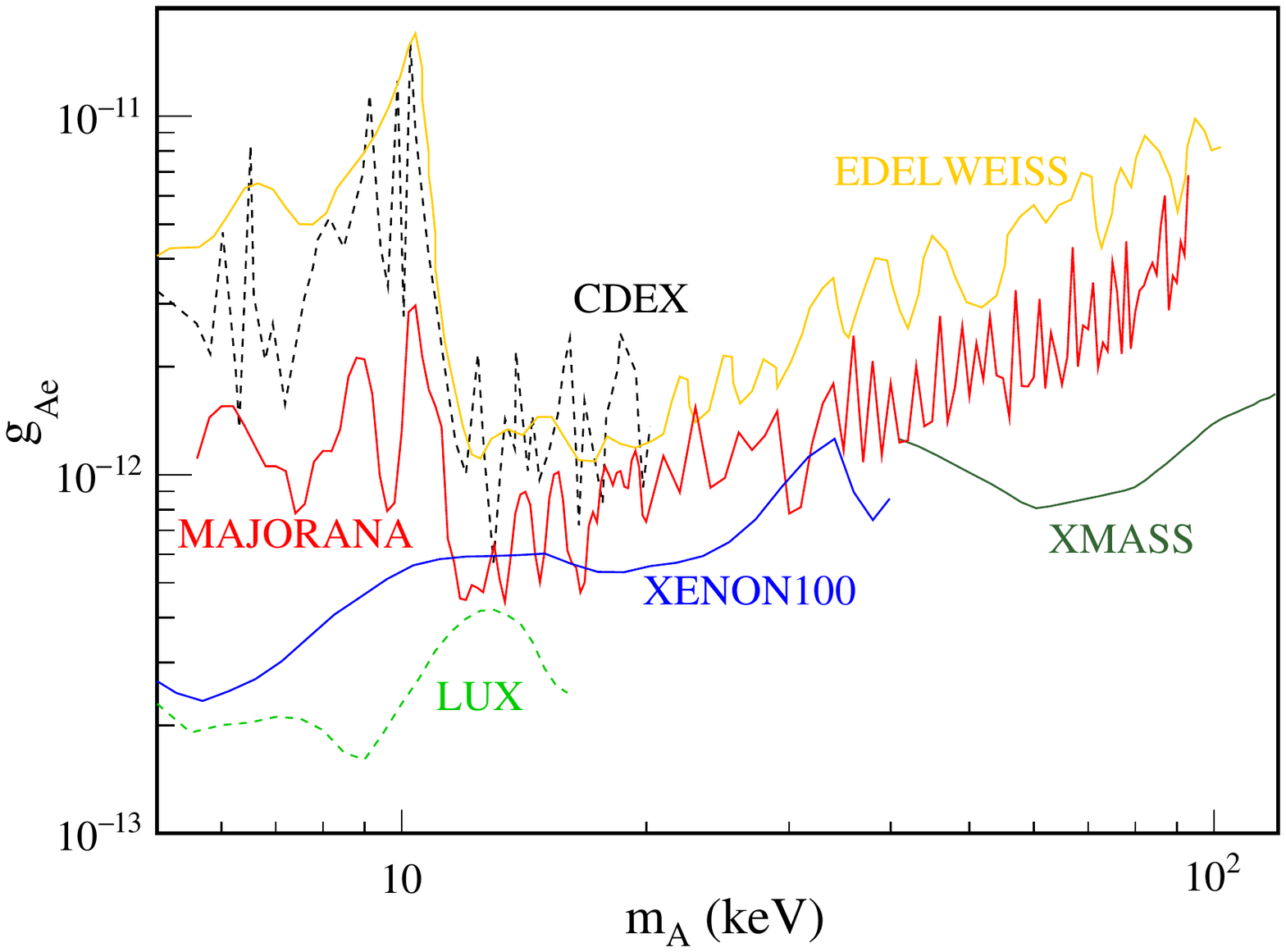} \\
    \end{array}$
    \caption{Left:  Energy spectra from the natural and enriched detectors in DS0 with 195~kg-d and 478~kg-d exposures respectively.  The dashed line shows a fit to a linear background with the tritium beta spectrum and the $^{68}$Ge K-shell peak.  For the natural detectors, tritium is dominant at low energy, and the $^{68}$Ge, $^{65}$Zn, and $^{55}$Fe peaks are indicated by arrows.  Right:  The 90\% upper limit on pseudoscalar dark matter from DS0 compared to other recent results (see~\cite{bosonicdm} for references).  \label{fig:ds0_lowe}}
  \end{center}
\end{figure}

The low background in the \dem, together with excellent energy resolution and low energy thresholds of PPC detectors, allows the \dem to have sensitivty to low energy searches for physics beyond the Standard Model.  Here we consider only low energy data from the DS0 commissioning which has a factor of 3-4 higher background than in later data sets since the inner UGEFCu shield had not been installed.  In the left panel of Figure~\ref{fig:ds0_lowe}, the DS0 low energy spectrum is shown for the natural and enriched detectors.  Control of the surface exposure of the enriched Ge has reulted in a dramatic reduction in low energy cosmogenic backgrounds, which are visible in the natural detector spectrum.  With the low backgrounds in the enriched detectors, the \dem can perform searches for pseudoscalar dark matter, vector dark matter, solar axions, and other exotic beyond the Standard Model physics.  Initial results from peak searches in the DS0 spectrum are presented in~\cite{bosonicdm} with 478~kg-d exposure, and the right panel of Figure~\ref{fig:ds0_lowe} shows the pseudoscalar dark matter limit obtained from the M1 commissioning data.  Improving instrumental background removal and analysis techniques for all datasets is ongoing, and will allow the analysis threshold to be lowered from 5~keV to near the sub-keV hardware trigger thresholds.

\section{Summary and outlook}
\label{sec:summary}

The \mjd began operating the first module containing enriched PPC detectors in-shield in June 2015, and both modules have been operating with the completed inner shield since August 2016.  The main goal of the \dem is to show that backgrounds can be reduced to the level which justifies a ton-scale \znbb experiment using \gei.  Initial backgrounds in the \znbb region of interest, estimated from operation of both modules in DS3 and DS4, are approaching the leading values set by the GERDA experiment~\cite{gerda}.  At the time of this writing, approximately 10 kg-y of enriched exposure has been acquired, and the analysis of the later datasets is being finalized in preparation for an initial \znbb result from the \dem.  With the increased exposure from later data sets and reduced analysis threshold, the \dem will also perform sensitive tests for beyond the Standard Model physics at low energy.

The recently formed LEGEND collaboration~\cite{legend} will combine the strengths of the \mjd and GERDA designs with the goal of a ton-scale \gei \znbb experiment capable of discovery level sensitivity in the inverted neutrino mass ordering region.

\ack
This material is based upon work supported by the U.S. Department of Energy, Office of Science, Office of Nuclear Physics, the Particle Astrophysics and Nuclear Physics Programs of the National Science Foundation, and the Sanford Underground Research Facility.

\end{document}